\documentstyle[12pt,aasms4]{article}
 
\slugcomment{To appear in the Astrophysical Journal (letters)
(accepted December 10, 1997)} 
 
\begin{document}
 
\title{An X-ray Cluster at  Redshift 2.156?}
 
\author{C.L. Carilli}
\affil{National Radio Astronomy Observatory, P.O. Box O, Socorro, NM,
87801 \\} 
\authoremail{ccarilli@nrao.edu}
\author{D.E. Harris}
\affil{Smithsonian Astronomical Observatory, 60 Garden St., Cambridge,
MA, 02138 \\} 
\author{L. Pentericci, H.J.A. R\"ottgering, G.K. Miley, and M.N. Bremer}
\affil{Leiden Observatory, Postbus 9513, 2300 RA Leiden, The
Netherlands \\}  

\begin{abstract}

We present X-ray observations of the narrow line radio galaxy 1138$-$262 at
z = 2.156 with the High Resolution Imager (HRI) on ROSAT.
Observations at other wave-bands, and in particular extremely
high values of Faraday rotation of the polarized radio emission, suggest 
that the 1138$-$262 radio source is in a dense environment,
perhaps a hot, cluster-type atmosphere. We detect 
X-ray emission from the vicinity of 1138$-$262, and we 
discuss possible origins for this
emission. The X-ray, optical, and radio data all
favor thermal emission from a hot cluster
atmosphere as the mechanism responsible for the X-rays, although we
cannot rule out a contribution from the active nucleus.
If this interpretation is correct, then 1138$-$262 
becomes the most distant, by far, of known X-ray emitting clusters. 
The X-ray luminosity for 1138$-$262 is
6.7$\pm$1.3 $\times$10$^{44}$ ergs 
sec$^{-1}$ for emitted energies between 2 keV and 10 keV.

{\sl Subject headings:} cosmology: large scale structure; galaxies: 
clusters -- active; radio continuum -- galaxies; X-rays -- galaxies

\end{abstract}

\section {Introduction}

Rich clusters of galaxies constitute the extreme high
mass end of the cosmic mass distribution function, and  study of such rare
objects provides unique leverage into theories  of structure
formation in the universe (Peebles 1983). 
Determining the cosmic
evolution of the X-ray luminosity function for the hot gaseous
atmospheres of  rich clusters can constrain both the power spectrum of
primordial density fluctuations, and the mean density of the
universe (Oukbir and Blanchard 1997, Oukbir, Bartlett, and Blanchard
1997, Henry 1997).  Unfortunately, optical studies of distant clusters can 
%Current data appear to favor a 
%universe with a mean density significantly less that the
%closure density, or world models with a finite cosmological  constant
be corrupted by field galaxy contamination (Frenk et al. 1990), while
X-ray detections of optically selected clusters have thus far been
limited to redshifts less than about one (Ebeling et al. 1997,
Castander et al. 1994, Nichol et al. 1997, Luppino and Gioia 1995,
Henry et al. 1997). 

One of the most effective methods for finding  galaxies, and possible
clusters,  at high
redshift is to target steep spectrum radio sources with faint optical
counter-parts (McCarthy 1993). Until recently the most
distant X-ray emitting cluster detected was that associated with the
radio galaxy 3C 295 at z = 0.461 (Henry and Henriksen 1986), and
the three most distant X-ray cluster candidates currently known
are all associated with powerful radio sources at z $\approx$ 
1 (Crawford and Fabian 1993, Crawford and Fabian 1995, Dickinson
1997). 

The radio source  1138$-$262 has been identified with a narrow
emission line galaxy at redshift 2.156 (Pentericci et al. 1997,
McCarthy et al. 1996). This radio galaxy is remarkable for a number of
reasons.  First, the optical continuum morphology is comprised of many
blue `knots' of emission, distributed over an area of  about 100 kpc
(Figure 1).\footnote{We adopt H$_o$ = 50 km sec$^{-1}$ Mpc$^{-1}$
and q$_o$ = 0.5.} 
The Ly$\alpha$ emission shows similar
clumpy structure on a similar scale.
Second, the radio source has the most disturbed morphology
of any z $>$ 2 radio galaxy yet identified,  indicating strong
interaction between the radio jet and the ambient medium.
And third  radio continuum polarimetric 
imaging of 1138$-$262 reveals the largest values of Faraday rotation 
for any radio galaxy at z $\ge$ 2 (Carilli et al. 1997),
with rotation measures as high as  6250 rad m$^{-2}$. 
Such large rotation measures are 
characteristic of nearby radio galaxies in dense cluster
atmospheres (Carilli et al. 1996, 
Taylor, Ge, and Barton 1994). 
All these data provide strong, although circumstantial,
evidence that the 1138$-$262 radio source is in a dense environment,
perhaps a hot, cluster-type atmosphere. 
To test this hypothesis we observed 1138$-$262 with the High
Resolution Imager on the ROSAT X-ray satellite.

\section{Observations, Results, and Analysis}

We observed 1138$-$262 with the  High
Resolution Imager on the ROSAT X-ray satellite (Tr\"umper 1983)
for 18 ksec on July 13, 1996, and for 23 ksec on June 11, 1997. 
The data were reduced using the IRAF-PROS data reduction software at
SAO, and the AIPS software at NRAO. 
%The X-ray photons were selected according to energy channel in
%order to reduce the background. 
The images from each epoch 
were registered using four `background' X-ray sources within
9$'$ of the target source. We estimate this relative registration is
accurate to about 1$''$. The  
absolute astrometry was determined  using the X-ray and
optical star, HD101551, located 9$'$ northwest of the target source. 
This astrometry was checked using two other X-ray-optical coincidences
in the field. From these we estimate a relative optical-to-X-ray
astrometric  accuracy of about 3$''$.

The ROSAT observations shows a clear
detection of an X-ray source close to the position of the radio and optical
nucleus of 1138$-$262 (Figure 2). The separation between the
X-ray peak and optical nucleus is 2$''$, which is smaller than 
the astrometric accuracy.
A total of 52 HRI counts were obtained in the optimum detection cell
of 21$''$ (see below), and the average  background in 
the source vicinity was 21 counts. Hence in the 41 ksec observation the
average net HRI count rate from 1138$-$262 was 0.75$\pm$0.16 counts
ksec$^{-1}$.

An important question concerning the X-ray emission from 1138$-$262 is
whether the emission is  spatially extended?
We have addressed this question in a number of ways using
the three brightest `background' sources within 6$'$ of 1138$-$262 
for comparison. 
First, we fitted an elliptical Gaussian function to the 
observed brightness distributions of 1138$-$262, and to the
background sources.
The results are summarized in Table 1. Columns 2 and 3 list the 
(deconvolved) Full Width at Half Maximum (FWHM) 
values, and the position angle of the major axis.
Second, we determined the FWHM
for the Gaussian smoothing function which 
optimizes the detection probability according to Poisson statistics. 
The FWHM values are listed in column 4 of Table 1, while column 5 lists
the net counts. In all cases the size parameters for 1138$-$262 are
larger than those of the background sources, although in the
case of the source 5.9$'$ to the east of 1138$-$262 the 
differences are within the errors.
And third, we calculated the average count rate per 1$''$ cell in
a series of concentric annuli centered on the position of each source.
The resulting radial profiles are shown in Figure 3.
Two of the sources appear to be marginally 
smaller than 1138$-$262, again
with the exception of the source 5.9$'$ to the east.
Overall, the values in the table, and the radial profiles,  suggest
that  the X-ray emission from 1138$-$262 may be more
extended than that from the  background sources, although in some
cases the differences are less than the errors.

%Given the paucity of counts from
%1138$-$262,  any definitive conclusions
%concerning the morphology of the X-ray emission remain problematic.

The HRI provides no information on the X-ray spectrum of
1138$-$262. In order to estimate the total  X-ray luminosity from
1138$-$262, we adopt  two different spectral models.
The first model is a thermal spectrum
with the same parameters as those determined for the X-ray emitting
cluster atmosphere enveloping the low redshift 
radio galaxy Cygnus A:~ kT = 7 keV and abundance = 30$\%$
solar (Reynolds and Fabian 1996, Ueno et al. 1994). Cygnus A was the first
extreme rotation measure radio source identified (Dreher et al. 1987),
and Cygnus A  has a similar radio luminosity to 1138$-$262.
The implied X-ray luminosity for 1138$-$262 is
6.7$\pm$1.3$\times$10$^{44}$ ergs 
sec$^{-1}$ for emitted energies between 2 keV and 10 keV.
For comparison, the luminosity for the Cygnus A
cluster in this band is 1.6$\times$10$^{45}$ ergs sec$^{-1}$.
The second model is a power-law with an energy index $-0.75$,
for which the  X-ray luminosity is 7.2$\pm$1.4$\times$10$^{44}$ ergs 
sec$^{-1}$.

Reynolds and Fabian (1996) present a detailed deprojection analysis of
the cluster X-ray emission from Cygnus A. They assume a cluster core
diameter of 300 kpc, and derive a density and pressure in the gas at the
core radius of 0.01 cm$^{-3}$ and 1.4$\times$10$^{-9}$ dyn cm$^{-2}$,
respectively, a total gravitational mass of 3$\times$10$^{14}$
M$_\odot$, and a total gas mass of 4$\times$10$^{13}$ M$_\odot$. 
For lack of better information we will assume in the analysis 
below that, if 1138$-$262 is an X-ray cluster, then  the 
Reynolds and Fabian numbers for
Cygnus A are representative of the 1138$-$262 environment.
The X-ray emission from the Cygnus A cluster  has a 
total extent of about 1 Mpc, with about half the emission coming from
the high surface brightness cluster core.
For 1138$-$262 the optimum detection  cell of 21$''$ corresponds to a 
physical scale of 170 kpc, comparable to the expected cluster core
diameter. It may be that, due to cosmological surface brightness
dimming, our HRI image of 1138$-$262 delineates 
only the higher surface brightness regions, while the low
surface brightness emission is lost in the background. 
In general, cosmological dimming will favor detection of clusters with
peaked X-ray surface brightness profiles at high redshift.

\section{Discussion}

Is the X-ray emission from 1138$-$262 from a hot cluster atmosphere, 
or from the active galactic nucleus (AGN)?  A critical criterion is
the spatial extent of the X-ray emission. Unfortunately,  given the
results on background sources in the field, and the paucity of photons
from 1138$-$262,  the current data is inconclusive on this
issue. 

There is much external, circumstantial,
evidence for a dense environment for 1138$-$262. First, the very distorted
radio morphology suggests a series of strong jet-cloud collisions
occurring over scales $\approx$ 100 kpc. 
Such interaction is supported by optical
spectroscopy, showing large velocities, and velocity dispersions, for
the line emitting gas in regions associated with the brightest radio
knots (Pentericci et al. 1997). Second, the optical morphology
consists of many blue knots of emission, with sizes between 5 and 10
kpc, again distributed over $\approx$ 100 kpc (Pentericci et
al. 1998). Pentericci et al. (1997) interpret this distribution as a group of
star forming galaxies in the act of merging into a giant elliptical
galaxy. This clumpy morphology is also
seen for the Ly $\alpha$ emitting gas, and the
major axis of the optical line and continuum emission is roughly
aligned with that of the radio source. 
And third, 1138$-$262 shows the most extreme
values of Faraday rotation for any galaxy at z $\ge$ 2 
(Carilli et al. 1997). Observations
of lower redshift radio galaxies have shown a  correlation between 
large rotation measures and cluster environment: all sources 
located at the centers of dense, X-ray emitting cluster atmospheres show
large amounts of Faraday rotation (Taylor et al. 1994). 
An important point is that this correlation is  independent
of radio source luminosity and morphological class, 
and hence is most likely a  probe of cluster
properties, and not radio source properties. The implication is that
the hot cluster gas must be substantially magnetized, with field strengths of
order a few $\mu$G (Carilli et al. 1996).

Worrall et al. (1994) have found a correlation between the radio and
X-ray emission from the nuclei of luminous, nucleus-dominated, radio
loud quasars, from which they conclude that the  dominant X-ray emission 
component is relativistically beamed synchrotron radiation (see also
Siebert et al. 1996). The radio
nucleus of 1138$-$262 has a flux density at 5 GHz  of:~ S$_{5 GHz}$ =
3.7 mJy. Applying the relationship from Worrall et al. then predicts
an X-ray flux density of:~ S$^{predicted}_{1 keV}$ =
4$\times$10$^{-4}$ $\mu$Jy. The observed value 
is a factor 40 larger:~ S$^{obs}_{1 keV}$ = 1.5$\times$10$^{-2}$ $\mu$Jy. 
%Moreover, the feature currently hypothesized as the radio
%nucleus in 1138$-$262 has a fairly steep spectral index, and hence it
%is possible that this feature is a knot in the jet, and that the true
%radio nucleus is even weaker than 3.7 mJy.
Worrall et al. also find that many lobe dominated quasars
fall above the nuclear X-ray-to-radio 
relationship defined by the nucleus-dominated
quasars  (in terms of X-ray flux density) 
by about the same factor as that seen for 1138$-$262, although the
scatter is large. They
hypothesize the existence of an extra (unknown) X-ray emission
component in some sources. In the case of 1138$-$262 we would argue
that this excess 
is likely to be thermal emission from a hot cluster atmosphere. 

The Worrall et al. relationships apply to broad emission line 
quasars, while 1138$-$262 is a narrow emission line galaxy. 
Perhaps a better comparison is made with the ultra-luminous 
narrow line radio galaxies Cygnus A and 3C 295, which have radio and
X-ray luminosities similar to 1138$-$262 (P$_{178 MHz}$ $\ge$
3x10$^{35}$ ergs sec$^{-1}$ Hz$^{-1}$, L$_x$ $\approx$ 10$^{45}$ ergs
sec$^{-1}$).  The non-thermal emission from the active nucleus of
Cygnus A  contributes at most 20$\%$ to the
total observed X-ray flux between 2 keV and 10 keV
(Ueno et al. 1994). 
For 3C 295, the point source contribution to the total observed
Einstein HRI flux is less than 13$\%$ (Henry and Henriksen 1986).
A similar result has been obtained
for the ultra-luminous radio galaxy 3C 324 at z = 1.2, which has
an associated X-ray emitting cluster with  size and luminosity
similar to 1138$-$262,  with no 
evidence of an X-ray  point source in the ROSAT HRI image (Dickinson
1997). It has been proposed that the
increased confinement of radio lobes 
provided by a dense local environment may enhance the conversion
efficiency of jet kinetic energy into radio luminosity,
thereby leading to a logical association between ultra-luminous radio
galaxies and dense cluster atmospheres (Barthel and Arnaud 1996).

One complication in the interpretation of 1138$-$262 which might argue
for a significant  AGN component is that
near infrared imaging of 1138$-$262 has revealed a
bright, small source coincident with the radio nucleus, with
K = 16 (Pentericci et al. 1997). This source is two magnitudes
brighter than  predicted by  the Hubble K-z relationship for
luminous radio galaxies (McCarthy 1993), perhaps suggesting that 
1138$-$262 is a  steep spectrum radio loud quasar in which the optical
nucleus is obscured by dust.  However, 
Pentericci et al. (1997) have shown that this
K band source  can be reasonably fit 
with a DeVaucouleur's law with a scale length of 2$''$ ($\approx$ 20
kpc), consistent with a bright elliptical galaxy, and 
significantly larger than the excellent seeing during the K band
observations (seeing FWHM $\approx$ 0.6$''$).
Pentericci et al. set a conservative upper
limit to the point source contribution in their image of 20$\%$. 
Unfortunately, their K band image may be contaminated by line emission
(H$\alpha$ and [NII]), which could bias the results depending on the 
relative distribution of the line and continuum emission.
%High resolution infrared imaging of 1138$-$262 in a line-free band  is
%required in order to address this issue.

The distorted jet-like morphology of the 
1138$-$262 radio source could also be construed as being more similar
to steep spectrum radio loud quasars than 
to most radio galaxies. Systematically more disturbed morphologies 
in quasars relative to radio galaxies are
thought to arise from a combination of projection 
(ie. jet axis not far from our line of sight),
and relativistic beaming (Barthel and Miley 1988).
A counter argument is that 1138$-$262 does
not have a prominent radio nucleus. The nuclear fraction for
1138$-$262 extrapolated to a rest frame frequency of 2.3 GHz is 
0.4$\%$ (assuming the core is flat spectrum below 5 GHz).
This fraction is comparable to values seen for  luminous
narrow line radio galaxies (typically $\approx$ 0.3$\%$),
but is considerably below values seen for broad line radio galaxies
(typically $\approx$ 3$\%$), and steep spectrum radio loud
quasars (typically $\approx$ 50$\%$), although the scatter in the
values for each type of source is large (Morganti et al. 1997).

Overall, the optical, radio, and X-ray data all favor 
a  hot cluster gas origin for the X-rays from
1138$-$262, however we cannot preclude an AGN contribution.
If  the emission is
dominated by the AGN, then the extreme rotation measures remain
unexplained, and 1138$-$262 would be very unusual among narrow line
radio galaxies in terms of the relative radio and X-ray properties
of its nucleus.

We proceed under the assumption that the X-ray emission from 
1138$-$262 is  from a hot cluster atmosphere at z = 2.156, and
we speculate on possible  physical implications. 
First, the over-density at which a structure is thought 
to separate from the general
expansion of the universe, $\rho_{sep}$, and collapse under its own
self-gravity is:~  $\rho_{sep}$
= ${9\pi^2}\over{16}$ $\times$ $\rho_{bg}$, where $\rho_{bg}$
is the mean cosmic mass density:~ $\rho_{bg}$ = 
7.5$\times$10$^{-29}$$\times$$\Omega$$\times$(1+z)$^3$ 
gm cm$^{-3}$ (Peebles 1993). 
At a redshift of 2.2 this becomes: $\rho_{sep}$(z=2.2) =
2.5$\times$10$^{-29}$$\times$$\Omega$ gm cm$^{-3}$.
The density of a typical X-ray cluster is:~ few$\times$10$^{-29}$ 
gm cm$^{-3}$ $\approx$ $\rho_{sep}$(z=2.2) for $\Omega$ = 1
(eg. mass $\approx$ 10$^{14}$ M$_\odot$, size $\approx$ 1 Mpc).
Hence, any large scale structure observed at z
$\ge$ 2 must be dynamically young, having just separated from the 
general expansion of the universe. Alternatively, massive structures
at large redshift might indicate  an open universe ($\Omega$ $<<$ 1).

Second,  the existence of a hot cluster-type atmosphere at high redshift
would argue for  shock heated,  primordial in-fall as the origin of
the cluster gas, rather than ejecta from cluster galaxies (Sarazin 1986). 
Such a model would require an early
epoch of star formation 
in order to supply the metals observed in the associated optical
nebulosity (Ostriker and Gnedin 1996). Third, the large rotation
measures observed  raise the question of the origin of  cluster
magnetic fields at early epochs. One possible mechanism is  
amplification of seed fields by a turbulent dynamo  operating 
during the formation process of the cluster as driven
by successive accretion events (Loeb and Mao 1994). The timescale for
field generation would then be a few times the cluster dynamical
timescale, or a few$\times$10$^{8}$ years.
Lastly,  these observations would then  
verify the technique of using extreme rotation
measure sources as targets for searches for cluster X-ray emission at high
redshift. Thus far there have been a total of 11 radio galaxies at
z $>$ 2 with source frame rotation measures greater than 1000 rad
m$^{-2}$, the most distant being  at z = 3.8 (Carilli et al. 1997,
Ramana et al. 1998, Carilli, Owen, and Harris 1994).

\vskip 0.2in

The National Radio Astronomy (NRAO) is a facility
of the National Science Foundation, operated under cooperative
agreement by Associated Universities, Inc..
D.E.H. acknowledges support from NASA contract NAS 5 30934.
We acknowledge support from a programme subsidy provided by the 
Dutch Organization for Scientific Research (NWO).

\newpage

\centerline{\bf References}

Barthel, P.D. and Arnaud, K. 1996,  M.N.R.A.S. (letters), 283, 45

Barthel, P.D., and Miley, G.K. 1988, Nature, 333, 319

Carilli, C.L., Owen, F., and Harris, D.E. 1994,  A.J., 107, 480

Carilli, C.L., R\"ottgering, H.J.A., van Ojik, R., Miley, G.K., and
van Breugel, W.J.M. 1997,  Ap.J. (Supp.), 109, 1

Carilli, C.L., Perley, R.A., R\"ottgering, H.J.A., and Miley, G.K. 1997,
in  IAU Symposium 175: Extragalactic Radio Sources, 
eds. C. Fanti and R. Ekers, (Kluwer: Dordrecht), p. 159

Castander, F., Ellis, R., Frenk, C., Dresseler, A., and Gunn,
J. 1994,  Ap.J. (letters), 424, L79

Crawford, C.S. and Fabian, A.C. 1993,  M.N.R.A.S. (letters), 260,
L15 

Crawford, C.S. and Fabian, A.C. 1995,  M.N.R.A.S., 273, 827

Dickinson, Mark 1997, in  {\sl 
HST and the High Redshift Universe}, eds. N. Tanvir, A. 
Aragon-Salamanca, and J.V. Wall, (World Scientific).

Dreher, J.W., Carilli, C.L., and Perley, R.A. 1987,  Ap.J., 316,
611

Ebeling, H., Edge, A.C., Fabian, A.C., Allen, S.W., Crawford,
C.S., and Bohringer, H. 1997,  Ap.J. (letters), 479, L101

%Fabbiano, G., Miller, L., Trinchieri, G., Longair, M., and Elvis,
%M. 1984,  Ap.J, 277, 115

Frenk, C.S., White, S.D., Efstathiou, G., and Davis, M. 1990, 
Ap.J., 351, 101

Henry, J.P. and Henriksen, M.J. 1986,  Ap.J., 301, 689

Henry, J.P. 1997,  Ap.J. (letters), 489, 1

Henry, J.P., Gioia, I.M., Mullis, C.R., Clowe, D.I., Luppino, G.A.,
Boehringer, H., Briel, U.G., Voges, W., and Huchra, J.P. 1997, A.J.,
114, 1293

Loeb, Abraham and Mao, Shude 1994,  Ap.J. (letters), 435, L109

Luppino, G.A. and Gioia, I.M. 1995,  Ap.J. (letters), 445,
L77 

McCarthy, P.J., Kapahi, V.K., van Breugel, Wil, Persson, S.E.,
Ramana, Athreya, and Subrahmanya, C.R. 1996,  Ap.J. (supp),
107, 19

McCarthy, P.J. 1993,  A.R.A.A., 31, 639

Morganti, R., Oosterloo, T.A., Reynolds, J.E., Tadhunter, C.N., and
Migenes, V. 1997, M.N.R.A.S., 284, 541

Nichol, R.C., Holden, B.P., Romer, A.K., Ulmer, M.P., Burke, D.J.,
and Collins, C.A. 1997,  Ap.J., 481, 644

Ostriker, Jeremiah P. and Gnedin, Nickolay Y. 1996, 
Ap.J. (letters), 472, L63

Oukbir, J. and Blanchard, A. 1997,  A\&A, 317, 1

Oukbir, J., Bartlett, J.G., and Blanchard, A. 1997,  A\&A,
320, 365 

Peebles, P.J.E. 1983,  {\sl The Large Scale Structure of the Universe},
(Princeton University Press: Princeton)

Peebles,  P.J.E. 1993,  {\sl Principles of Physical Cosmology},
(Princeton Univ. Press: Princton)

Pentericci, L., R\"ottgering, H.J.A.,  Miley, G.K., Carilli, C.L., and
McCarthy, P. 1997,  A\&A, 326, 580

Pentericci, L., R\"ottgering, H.J.A.,  Miley, G.K., and
McCarthy, P. 1998,  Ap. J. (letters), submitted

Ramana, A., Kapahi, V.K., McCarthy, P., and Subramaniah, R. 1998, 
 Ap.J., submitted.

Reynolds, C.S. and Fabian, A.C 1996,  M.N.R.A.S., 278, 479

R\"ottgering, H.J.A. 1993,  Ph.D. Thesis, University of Leiden

Sarazin, Craig L. 1986,  Rev. Mod. Phys., 58, 1

Siebert, J., Brinkmann, W., Morganti, R., Tadhunter, C.N., Danziger,
I.J., Fosbury, R.A.E., and di Serego Alighieri, S. 1996, M.N.R.A.S.,
279, 1331 

Stark, A.A., Gammie, C.F., Wilson, R.W., Bally, John, Limke, Richard,
Heiles, Carl, and Hurwitz, M. 1992,  Ap.J. (Supp.), 79, 77

Taylor, G.B., Ge, J.-P., and Barton, E. 1994,  A.J., 107,
1942

Tr\"umper, J. 1983, Adv. Space Res., 2, 241 

Ueno, Shiro, Koyama, Katsuji, Nishida, Minoru, Shigeo, Yamauchi, and
Ward, Martin J. 1994,  Ap.J. (letters), 431, L1

Worrall, D.M., Lawrence, C.R., Pearson, T.J., and Readhead,
A.C.S. 1994,  Ap.J. (letters), 420, L17

\clearpage
\newpage

\begin{deluxetable}{ccccc}
\footnotesize
\tablecaption{Source Size Parameters \label{tbl-1}}
\tablewidth{0pt}
\tablehead{
\colhead{Source} & \colhead{FWHM}  & 
\colhead{PA} & \colhead{Optimum Cell} & \colhead{Net Counts} \nl
\colhead{~} & \colhead{arcsec}   & \colhead{deg} 
& \colhead{arcsec~} & \colhead{~} \nl
}
\startdata
1138-262 & 14$\times$9 $\pm$ 2 & 52 $\pm$ 11 & 
21 $\pm$ 2  & 31 $\pm$ 7 \nl
5.8$'$ West & 9$\times$7 $\pm$ 1 & 107 $\pm$ 10 & 15 $\pm$ 2 &
38 $\pm$ 7 \nl
5.9$'$ East &  13$\times$6 $\pm$ 1 & 143 $\pm$ 11 &  19 $\pm$ 2 & 28
$\pm$ 7\nl
3.8$'$ Southeast &  10$\times$6 $\pm$ 2 & 153 $\pm$ 15 &13 $\pm$ 2 &
16 $\pm$ 5 \nl 
\enddata
\end{deluxetable}

\clearpage
\newpage

\centerline{Figure Captions}

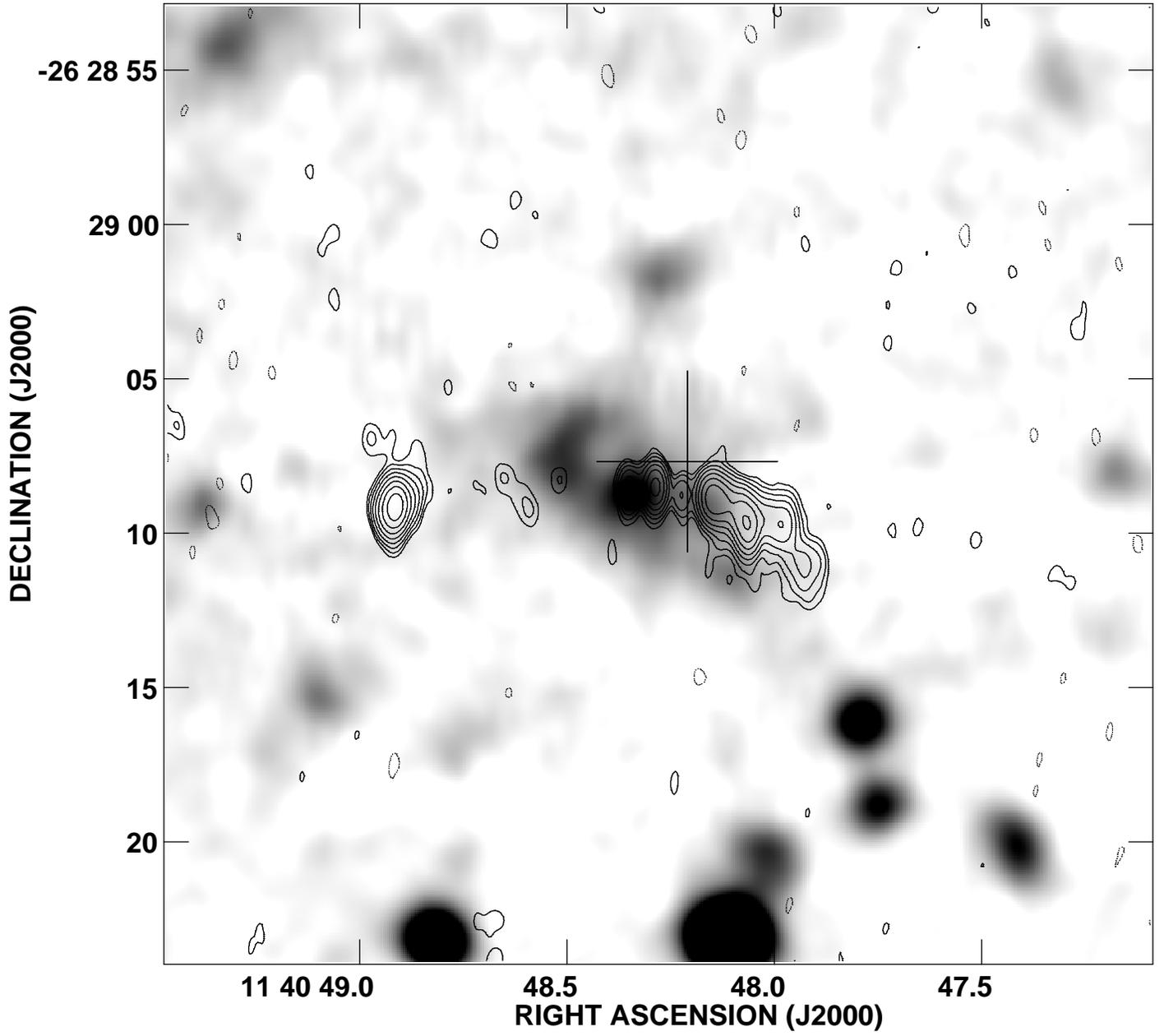
\noindent Figure 1 -- Contours show the Very Large Array 
radio image of 1138$-$262 at 5 GHz (Carilli et al. 1997) with 
a resolution of 0.4$''$$\times$0.7$''$. 
The contour levels are a geometric
progression in square root two, with the first level being 
0.15 mJy beam$^{-1}$.
The grey-scale shows the optical image (R + B bands) of
1138$-$262 from the New Technology Telescope (Pentericci et al. 1997). 
The cross indicates the position of X-ray peak surface brightness, and 
the cross size indicates the estimated astrometric accuracy. 

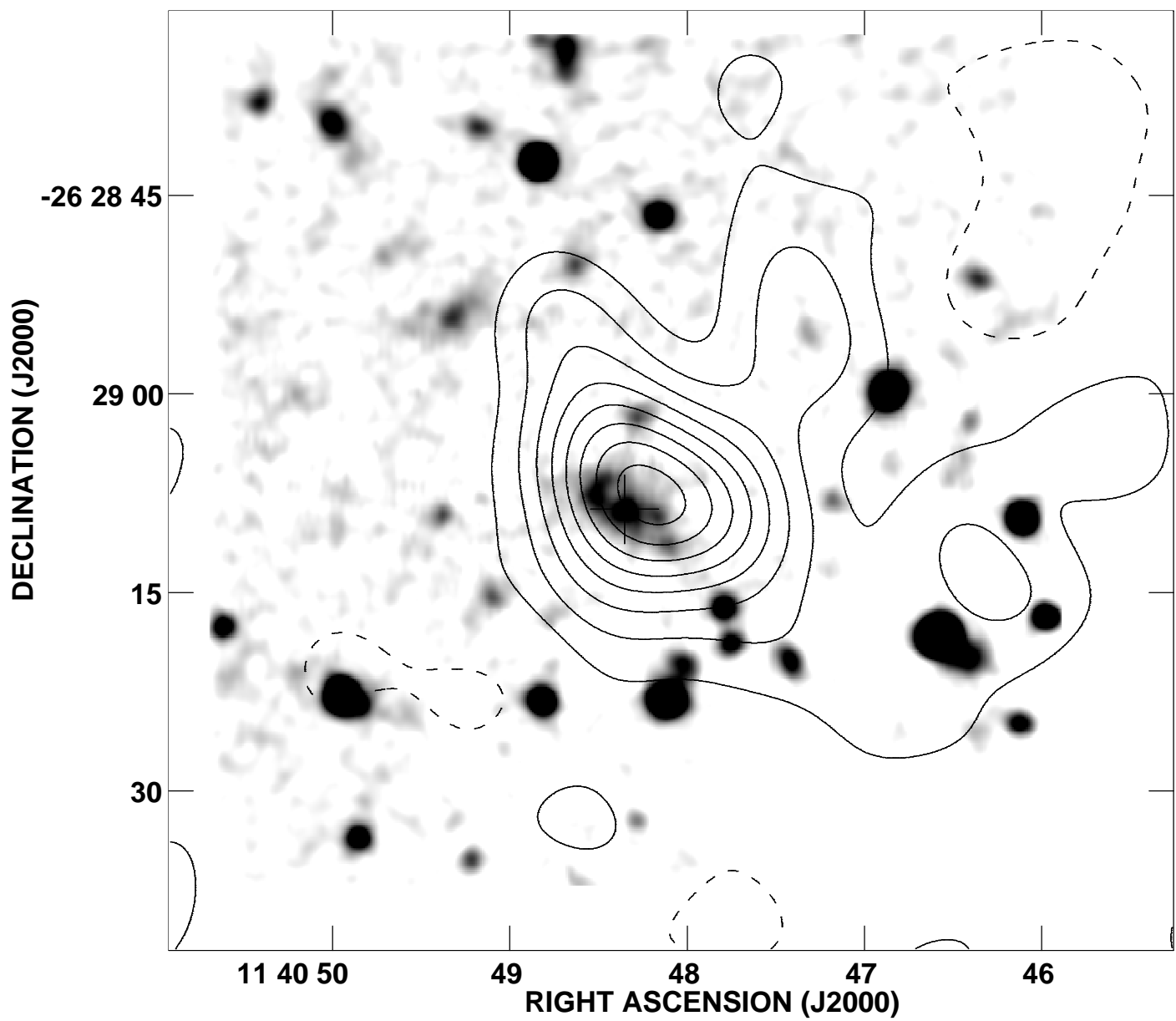
\noindent Figure 2 -- The X-ray image of   1138$-$262 from a
41 ksec exposure with the ROSAT HRI.
The image has been convolved with a Gaussian beam of FWHM = 10$''$,
and a   mean background level of 5 counts per beam has been subtracted. 
The contour levels are: -4, -2, 2 4 6 8 10 12 14 16 counts per beam.
The cross shows the position of the radio and optical nucleus
of 1138$-$262, and the cross size indicates the estimated astrometric
accuracy.  For the thermal spectral model discussed
in section 2, two counts in 41 ksec implies
an X-ray flux of 1.9x10$^{-15}$ ergs cm$^{-2}$
sec$^{-1}$ in the emitted 2 - 10 keV band, using 
the Galactic HI column density of 4.5$\times$10$^{20}$ cm$^{-2}$
(Stark et al. 1992). 

\noindent Figure 3 -- The curves show the mean counts per 1$''$ cell
in a series of concentric annuli centered on the position of
1138$-$262 (solid line), and on the positions of the three
brightest X-ray sources in the field located within 6$'$ of
1138$-$262. All the curves have been normalized to a peak of unity,
and the mean background of 0.05 counts per cell has been subtracted.
The dotted line is for the source 5.8$'$ to the west of 1138$-$262,
the short dashed line is for the source 5.9$'$ to east, and the 
long dashed line is for the source 3.8$'$ to the southeast.
Each annulus has a width of 3$''$. The 
error bars are based on photon counting statistics for the target 
source. 

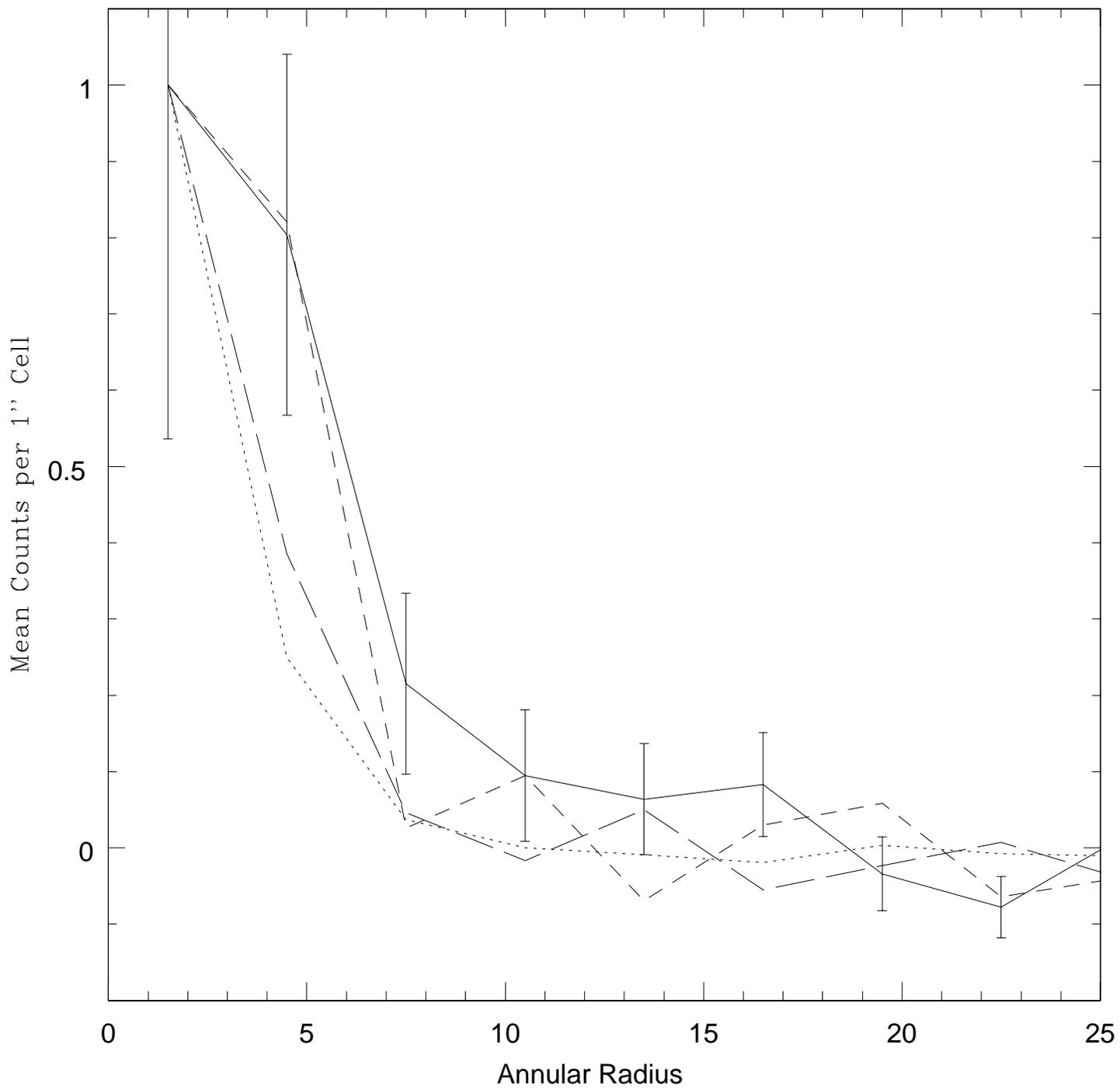
\end{document}